\documentclass[conference]{IEEEtran}
\IEEEoverridecommandlockouts
\usepackage{cite}
\usepackage{amsmath,amssymb,amsfonts,amsthm}
\usepackage[linesnumbered,lined]{algorithm2e}
\usepackage{graphicx}
\usepackage{textcomp}
\usepackage{xcolor}
\def\BibTeX{{\rm B\kern-.05em{\sc i\kern-.025em b}\kern-.08em
    T\kern-.1667em\lower.7ex\hbox{E}\kern-.125emX}}
    
\theoremstyle{definition}

\newtheoremstyle{theorem}%		% Name
	{}%							% Space above
	{}%							% Space below
	{\itshape}%							% Body font
	{}%							% Indent amount
	{\bfseries}%							% Theorem head font
	{.}%							% Punctuation after theorem head
	{ }%							% Space after theorem head, ' ', or \newline
	{}%							% Theorem head spec (can be left empty, meaning `normal')
\theoremstyle{theorem}

\allowdisplaybreaks

\begin{document}

\title{Cyber Orbits of Large Scale Network Traffic
\thanks{Research was sponsored by the Department of the Air Force Artificial Intelligence Accelerator and was accomplished under Cooperative Agreement Number FA8750-19-2-1000. The views and conclusions contained in this document are those of the authors and should not be interpreted as representing the official policies, either expressed or implied, of the Department of the Air Force or the U.S. Government. The U.S. Government is authorized to reproduce and distribute reprints for Government purposes notwithstanding any copyright notation herein.
Use of this work is controlled by the human-to-human license listed in Exhibit 3 of https://doi.org/10.48550/arXiv.2306.09267}
}

\author{\IEEEauthorblockN{Jeremy Kepner, Hayden Jananthan, Chasen Milner, Michael Houle, \\ Michael Jones, Peter Michaleas, Alex Pentland
\\
\IEEEauthorblockA{MIT
}}}

\maketitle

\begin{abstract}
	The advent of high-performance graph libraries, such as the GraphBLAS, has enabled the analysis of massive network data sets and revealed new models for their behavior.  Physical analogies for complicated network behavior can be a useful aid to understanding these newly discovered network phenomena.  Prior work leveraged the canonical Gull’s Lighthouse problem and developed a computational heuristic for modeling large scale network traffic using this model.  A general solution using this approach requires overcoming the essential mathematical singularities in the resulting differential equations.  Further investigation reveals a simpler physical interpretation that alleviates the need for solving challenging differential equations.  Specifically, that the probability of observing a source at a temporal ``distance'' $r(t)$ at time $t$ is $p(t) \propto 1/r(t)^2$. This analogy aligns with many physical phenomena and can be a rich source of intuition.  Applying this physical analogy to the observed source correlations in the Anonymized Network Sensing Graph Challenge data leads to an elegant cyber orbit analogy that may assist with the understanding network behavior.
\end{abstract}


\section{Introduction}

The high performance GraphBLAS math library \cite{davis2019algorithm} has enabled the holistic analysis of network traffic events from the worlds' largest Internet telescope (CAIDA \cite{caida2023ucsd}) and commercial honeyfarm (GreyNoise \cite{greynoise2023greynoise}).  These analysis \cite{kepner2022temporal, jananthan2023mapping, kepner2024normal} indicate that the probability $p(t)$ of an observed Internet source being seen again at time $t$ follows a modified Cauchy distribution
\begin{equation}
	p(t) \propto \frac{1}{t_{0.5}^\alpha + t^\alpha}
\end{equation}
with the exponent $ 0 < \alpha < 1$ and $t_{0.5} > 0$ (the time at which probability decreases to 0.5). When $\alpha = 2$, the function becomes the standard Cauchy distribution, a canonical example of a heavy-tailed distribution \cite{nair2022fundamentals}, for which a simple geometric interpretation is given by Gull’s Lighthouse problem \cite{gull1988bayesian}.  A general solution for Gull’s Lighthouse problem for diverse geometries requires overcoming the essential mathematical singularities in the resulting differential equations and prior work explored various heuristic approaches to overcoming these challenges \cite{jananthan2023mapping}.

  Another approach is to reassess the underlying physical interpretation of the Gull’s Lighthouse problem whose motivation was to show students how application of Bayes' theorem could produce the Cauchy distribution \cite{gull1988bayesian}.  Interestingly, a far simpler physical interpretation exists that leads to the same result without  intermediate steps.  Specifically, the probability of observing a source at a temporal ``distance'' $r(t)$ is $p(t) \propto 1/r(t)^2$. This inverse-squared distance analogy aligns with many physical phenomena (e.g., gravitation, charged particles, sound, light, and other electromagnetic radiation) and leads to the direct translation of observed Internet source correlation probabilities into a ``cyber orbit"
\begin{equation}
	r(t) \propto \frac{1}{\sqrt{p(t)}}
\end{equation}
where the units of distance are measured in time.  Completing the orbital interpretation also requires defining an angle $\theta(t)$.  Various functions can be used to translate the time dimension into an angle. Perhaps the simplest is to derive the angle from a triangle with base $\Delta t$ and height $r(t)$
\begin{equation}
	\Delta \theta(t) \approx \sin(\Delta \theta(t))  = \frac{\Delta t}{r(t)}  ~~~~~~~ {\rm [see ~ Fig.~1]}
\end{equation}
for small values of $\Delta \theta(t)$. The corresponding cyber orbits can then be understood in terms of potentially more accessible concepts akin to high-flying satellites and low-flying rockets.

To illustrate these concepts the above cyber orbital equations are applied to CAIDA and GreyNoise data from the Anonymized Network Sensing Graph Challenge \cite{jananthan2024anonymized}.

\section{Data Set}

CAIDA collects over 1,000,000,000,000 unique packets each month from hundreds of millions of unique sources. GreyNoise converses with millions of unique sources each month. This volume of data requires advanced technologies to analyze, such as, supercomputers (e.g., the MIT SuperCloud \cite{reuther2018interactive}) and high performance math libraries (e.g., the GraphBLAS \cite{davis2019algorithm}).  These technologies have made it possible to analyze hundreds of billions of packets in minutes and played key roles in providing data for the Anonymized Network Sensing Graph Challenge \cite{jananthan2024anonymized}.  The Graph Challenge has released a number of representative CAIDA telescope and GreyNoise honeyfarm data sets that are amenable to cyber orbit analysis.  

\begin{figure*}[htbp]
	\begin{center}
		\includegraphics[width=\columnwidth]{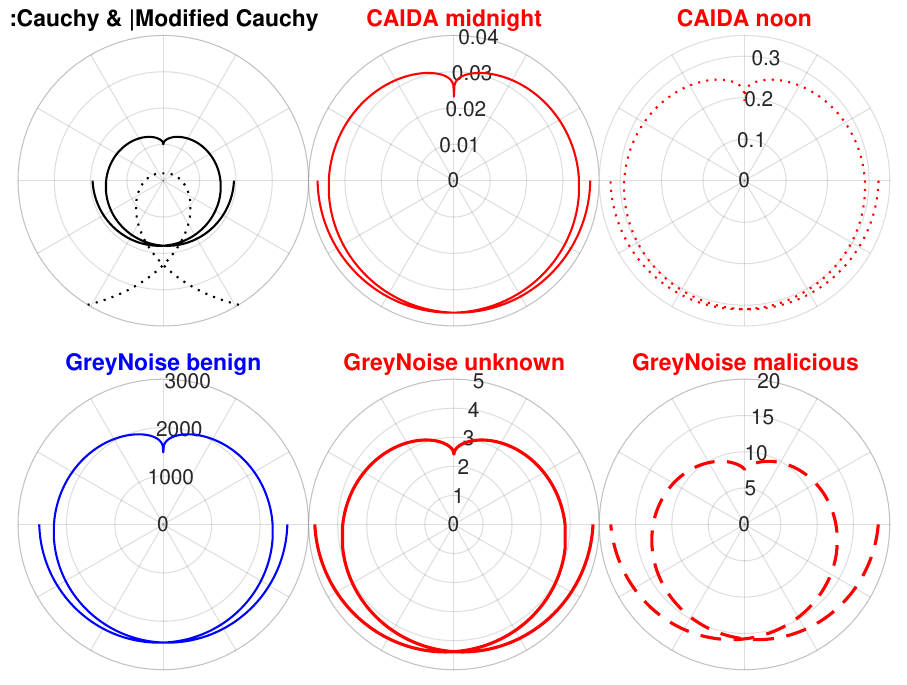}
		\includegraphics[width=\columnwidth]{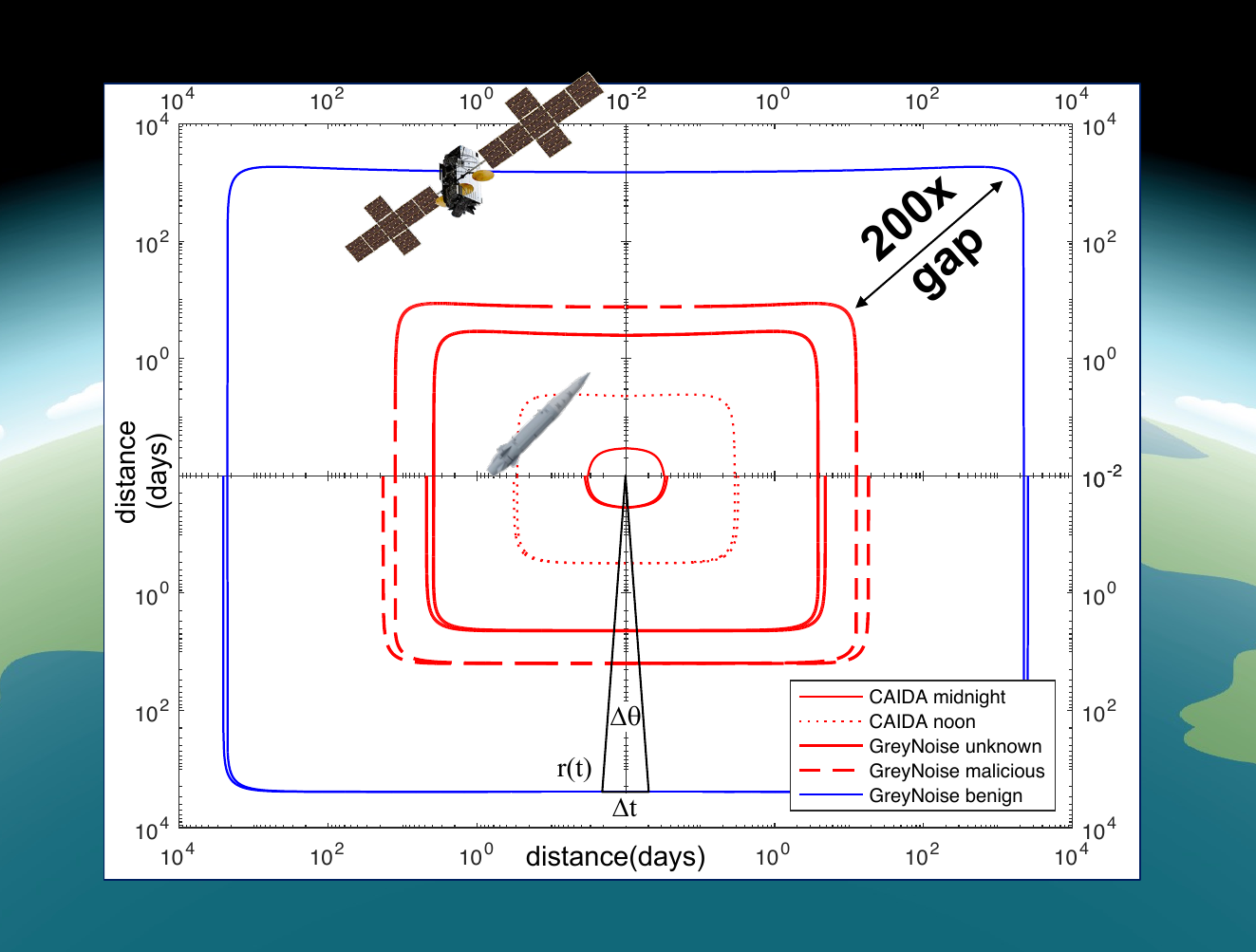}
	\end{center}
	\caption{(left) Polar plots of the cyber orbits.  Upper left shows representative orbits for Cauchy (black dots) and modified Cauchy (black line) distributions.  Peaks in the probability correspond to decreases in the orbits.  The long term behavior of the CAIDA telescope and GreyNoise honeyfarm orbits are slowly increasing orbits. (right) Logarithmic Cartesian plots of the CAIDA telescope and GreyNoise honeyfarm orbits showing the 200x gap between benign and malicious orbits akin to high flying satellites and low-flying rockets.  Geometric construction of equation (3) is depicted with the black triangle.}
\label{cyberorbits}
\end{figure*}

The CAIDA telescope data consists of several hundred $2^{30}$ packet collections of anonymized sources and destinations collected at noon and midnight over several months.  The GreyNoise data consists of one month samples of anonymized sources collected over the course of a year.  The CAIDA telescope observes an unassigned ``darknet'' and nearly all the traffic can be ascribed to malicious activities (e.g., scanning and DDoS backscatter).  The GreyNoise honeyfarm engages in active conversations with Internet sources and is able to categorize these conversations as benign, malicious, or unknown (likely maliciously).

The temporal source correlations and corresponding best-fit modified Cauchy distributions $p(t)$ are computed as described in \cite{jananthan2023mapping}.  The cyber orbits $r(t)$ and $\theta(t)$ are computed from $p(t)$ using equations (2) and (3).

\section{Results \& Conclusions}

The computed cyber orbits are shown in Figure~\ref{cyberorbits} as polar plots (left) and logarithmic Cartesian plots (right).  The polar plots illustrate how the peaks in probability correspond to decreases in the orbits at $\theta(t=0) =0$.  Likewise, the long term behavior of the CAIDA telescope and GreyNoise honeyfarm orbits are slowly increasing orbits $\propto \sqrt{t_{0.5}^\alpha + t^\alpha}$. The logarithmic cartesian plots of the CAIDA telescope and GreyNoise honeyfarm orbits highlight the 200x gap between the benign and malicious orbits akin to high flying satellites and low-flying rockets.  Such gaps, have the potential for aiding in discerning benign from malicious traffic.

This physical analogy for the complex network behavior of benign and malicious Internet sources is offered up as a potentially useful aid to understanding these network phenomena.  Visualizing the transient behavior of Internet communications as orbits opens new possibilities for considering how to reason about and explain these phenomena a quasi-periodic systems.  More specifically, Internet sources mostly maintain a relatively slow changing distance between each other, but occasionally move closer when their interactions increase.  For malicious sources with small $t_{0.5}$ this gap closes and opens quickly, while for benign sources with large $t_{0.5}$ the gap closes and opens slowly.
  
\section*{Acknowledgment}

The authors wish to acknowledge the contributions and support of:  Andersen, Anderson, Arcand, Atkins,  Bergeron, Birardi, Bond, Bonn, Byun, Burrill, Claffy, Davis, Demchak, Edelman, Fisher, Gadepally, Gottschalk,  Grant,  Hardjono, Hill, Houle, Hubbell, Leiserson,  Luszczek, Malvey, Milechin, Mohindra,  Morales, Morris, Mullen, Patel, Perry, Pisharody, Prothmann,  Prout, Rejto,  Reuther, Rosa, Ruppel,  Rus,  Sherman, Somin, Wachman, Weed,  Yee, Zissman.

%\section*{References}

\bibliographystyle{IEEEtran}
\bibliography{CyberOrbits}

\end{document}